\documentclass{elsart41}
\usepackage{graphics}
\usepackage{graphicx}
\usepackage{amssymb}
\begin{document}

\begin{frontmatter}



\title{Possible Localization Behavior of the Inherent
Conducting Polymer (CH$_3$)$_{0.9}$ReO$_3$}


\author[CPM]{E.--W. Scheidt\corauthref{Scheidt}},
\ead{Scheidt@physik.uni-augsburg.de}
\author[CPM]{R. Miller},
\author[CPM]{Ch. Helbig},
\author[CPM]{G. Eickerling},
\author[CPM]{F. Mayr},
\author[CPM]{R.~Herrmann},
%\author[CPM]{K. Tr\"oster},
\author[TP]{P. Schwab},
\author[CPM]{W. Scherer}

\address[CPM]{Chemische Physik und Materialwissenschaften,
Universit\"{a}t Augsburg, 86159 Augsburg, Germany}
\address[TP]{Theoretische Physik II,
Universit\"{a}t Augsburg, 86159 Augsburg, Germany}

\corauth[Scheidt]{E.--W. Scheidt. Tel: +49\,821\,5983356 fax:
+49\,821\,5983227}

\begin{abstract}

Polymeric methyltrioxorhenium (poly-MTO) represents the first
example of an inherent conducting organometallic oxide. It adopts
the structural motives and transport properties of some classical
perovskites in two dimensions. In this study we present
resistivity  data down to 30\,mK which exhibit a crossover from a
metallic  (d$\rho$/d$T >$ 0) to an insulating (d$\rho$/d$T <$ 0)
behavior at about 30\,K. Below 30\,K an unusual resistivity
behavior, similar to that of some doped cuprate systems, is
observed: initially the resistivity increases approximately as
$\rho \sim$ log$(1/T$) before it starts to saturate below 2\,K.
Furthermore, a linear positive magnetoresistance is found (up to
7\,T). Temperature dependent magnetization and specific heat
measurements in various magnetic fields indicate that the unusual
resistivity behavior may be driven by spatial localization of the
d$^1$ moments at the Re atoms.

\end{abstract}

\end{frontmatter}

% main text

Polymeric methyltrioxorhenium (poly-MTO),
\{(CH$_{3}$)$_{0.9}$ReO$_{3}$\}$_{\infty}$, is a unique
representant of a conductive organometallic polymer in metal-oxide
systems, with a moderate high resistivity of 6\,m$\Omega$cm at
room temperature \cite{Hermann_1995}.
%\cite{Hermann_1992}
The conductivity is attributed to a fraction of demethylated Re
atoms. Instead of a crossover from a Re(VII) (d$^0$) to a Re(VI)
(d$^1$) state, these demethylated Re atoms are effectively
oxidized and their electrons are transfered to the band system.
%Therefore the net effect of a demethylation of a few
%Re atoms creates an effective reduction of each Re atom.
Only a minor part (0.05$\%$ Re atoms \cite{Miller_2005}) remains
located at the metal sites which are in the following treated as
Re(d$^1$) centers. They model a two dimensional dilute metal-oxide
spin system. The attempt to increase the electronic conductivity
of poly--MTO by employing the organic donor species
tetrathiafulvalene (TTF) leads to  a crossover from metallic to
insulating behavior with increasing TTF contribution
\cite{Miller_2005}.

The resistivity of poly--MTO  at low temperatures and in high
magnetic fields within the ReO$_2$ planes resembles $\rho$ of the
CuO$_2$ planes of the Zn--doped high--$T_\mathrm{c}$
superconductor YBa$_{2}$Cu${}_3$O$_{7-\delta}$ (YBCO)
\cite{Segawa_1999}. The scattering centers in Zn--doped YBCO are
due to nonmagnetic Zn centers in the antiferromagnetic spin
correlated CuO$_2$ planes. The scattering centers of poly--MTO
mirror the inverse situation: the magnetic d$^1$ centers are
placed in nonmagnetic ReO$_2$ planes. Therefore poly--MTO might be
a promising candidate to revitalize the discussion about the
electron scattering mechanism in cuprates.

%####################################################################
\begin{figure}[!ht]
\begin{center}
\includegraphics[width=0.45\textwidth]{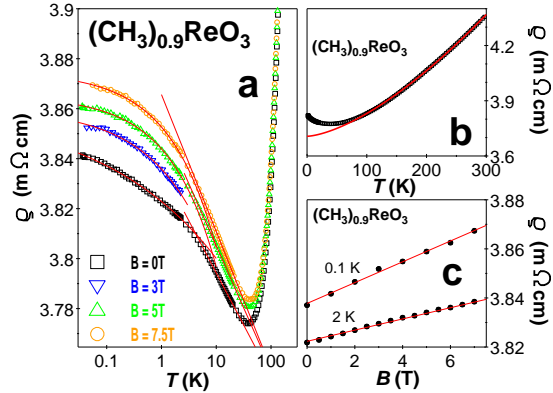}
\end{center}
\caption{(a) Resistivity vs. log$T$ in various magnetic fields
$B$. The solid lines are logarithmic fits between 5\,K and 30\,K
and power--law fits below 2.5\,K  ($\Delta \rho \propto
T^{\alpha}$ with $\alpha = 0.2; 0.4; 0.5; 0.5$ for $B = 0; 3; 5;
7$\,T, respectively). (b) Above 80\,K the resistivity obeys a
power--law $\Delta \rho \propto T^{1.5}$. (c) A positive
magnetoresistivity is observed, similar to that seen in
Ref.~\cite{Segawa_1999}.} \label{fig1}
\end{figure}
%####################################################################

Poly-MTO was synthesised by  auto-polymerization of MTO in flux at
120\,$^\circ$C during 48\,h  \cite{Miller_2005}. X-ray powder
diffraction measurements suggest a two-dimensional
\{ReO$_{2}$\}$_{\infty}$ layered structure. The missing 00\emph{l}
series and the asymmetric shape of the \emph{hk}0 reflections
indicate ordering to occur solely in two dimensions.

Figs.~\ref{fig1}\,a,b show the resistivity of
(CH$_{3}$)$_{0.9}$ReO$_{3}$ on a semi--logarithmic and linear
temperature scale, respectively. The high residual resistivity may
be due to interlayer disorder \cite{Miller_2005}. The temperature
dependence of the resistivity clearly exhibits a crossover from a
metallic (d$\rho$/d$T >$ 0) to an insulating (d$\rho$/d$T <$ 0)
behavior at about 30\,K. Below 30\,K a log$(1/T$) divergence over
one decade of $T$ is observed, similar to that found in Zn--doped
YBCO \cite{Segawa_1999}. At lowest temperatures a crossover to a
power--law dependence is detected, $\Delta \rho \propto
T^{\alpha}$, with $\alpha \approx 0.5$ for $B
> 3$T. In the insulating regime we observe a positive, linearly
increasing magnetoresistivity as depicted for two temperatures in
Fig.~\ref{fig1}\,c.

Common scenarios predicting a log$(1/T)$ behavior like
conventional Kondo impurities or 2D weak localization, cannot
explain our experimental results. In both cases the logarithmic
divergence of the resistivity should be reduced in the presence of
a magnetic field.

\begin{figure}[!ht]
\begin{center}
\includegraphics[width=0.43\textwidth]{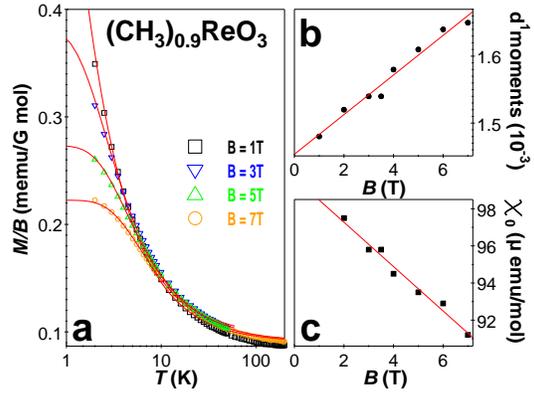}
\end{center}
\caption{(a) $M/B$ of poly--MTO in several external magnetic
fields. The solid lines are Brillouin-fits for a d$^1$ moment
using the amount of Re(d$^1$) centers (b) and $\chi_0$ (c) as fit
parameters.} \label{fig2}
\end{figure}

For  further information concerning the origin of the positive
linear magnetoresitivity, temperature dependent magnetization
measurements in various magnetic fields $B$ were performed. In
Fig.~\ref{fig2}\,a the magnetization $M$ divided by  the applied
magnetic field $B$ is plotted vs. temperature. The solid lines are
fits, which follow a Brillouin function, assuming that the
paramagnetic behavior is only due to independent d$^1$ moments
with a quenched orbital moment. The two fit parameters, the amount
of Re(d$^1$) centers and the constant itinerant contribution
$\chi_0 = \chi_{\mathrm{Pauli}}+ \chi_{\mathrm{Landau}}$, are
pictured in Figs.~\ref{fig2}\,b and c, respectively. These two
plots show clear evidence, that with increasing magnetic field,
the amount of localized d$^1$ moments increases linearly with a
simultaneous decrease of the itinerant electrons. This spatial
localization of the d$^1$ moments at the Re atoms might be the
origin of the unusual linear positive magnetoresitivity in
poly--MTO.

%####################################################################
%####################################################################

This interpretation is also corroborated by the magnetic field
dependence of the internal electric field gradient ($V_{zz}$) at
the Re site. Analysis of specific heat data  below 1\,K with a
crystal field model reveals a decrease of the Re nuclear
quadrupole splitting  with increasing magnetic field $B$, pointing
to a decrease of $V_{zz}$. This indicates a reduction of the
electronic and structural anisotropy at the Re site, which is in
good agreement with density functional theory (DFT) geometry
optimization of poly--MTO, where an increasing amount of localized
d$^1$ centers leads to a reduction of strain in the ReO$_2$ planes
\cite{Eickerling_2005}.

The origin of the log$(1/T)$ and  $\sqrt T$ dependence of $\rho$
could be the Altshuler--Aronov \cite{Altshuler-Aronov 85}
correction in the presence of a crossover from 2D to 3D diffusion
at lower temperature. In a granular system a similar crossover is
also expected between the high temperature incoherent tunnelling
and the low temperature coherent intergrain tunnelling
\cite{Beloborodov04}. But at present we cannot exclude Kondo--like
scenarios. In this respect we notice that the spatial localization
of the d$^1$ moments in the ReO$_2$ planes might be a new approach
for the understanding of the unusual resistivity of Zn-doped YBCO.

%Research on highly correlated electron systems within the
%preceding thirty years has fascinated many of us \cite{Name1}.
%The novel binary compound
%$\rm CeCu_{7.77}$, with ferromagnetic ordering at $T = 10$~K and
%a transition into a superconducting ground state below $T_c = 5$~K, however,
%was the highlight of the previous week \cite{Name2}. \dots
%\vspace{0.5cm}
%If you like to safe space in your
%two (four) page contribution to the
%SCES'05, we recommend not to use section headings like
%\vspace{0.3cm}
%A. Name1, et al., Physica {\bf B} 500 (2008) 333.
%\vspace{0.3cm}
%Please provide your figures as EPS file; they will then automatically
%match the available space when using the commands
%as indicated in the next few lines. fig1 is YOUR figure 1.
%
%
%\begin{figure}[!ht]
%\begin{center}
%\includegraphics[width=0.45\textwidth]{Figure_1.eps}
%\end{center}
%\caption{This is the first picture}
%\label{fig1}
%\end{figure}
%
%
%An easy
%way to obtain appropriate EPS files is to virtually print your
%figure using Adobe Acrobat as ``printer'' and then convert the PDF file
%into an EPS file [normally this is done by "save as" and then select the
%``save as type'' Encapsulated PostScript (*.eps), e.g., Acrobat 6.0]
%Standard TeX and LaTeX commands fully agree with the present
%style file of Elsevier.

\section*{Acknowledgement}
This work was supported by the SFB~484 of the Deutsche
Forschungsgemeinschaft (DFG).

\end{document}